\documentclass[aps,prc,twocolumn,superscriptaddress]{revtex4-1}

\usepackage{verbatim}
\usepackage{multirow}
\usepackage{amsmath}
\usepackage{amssymb}
\usepackage{siunitx}
\usepackage{units}
\usepackage{graphicx}
\usepackage{adjustbox}
\usepackage{float}
\usepackage{dcolumn}
\usepackage{physics}
\usepackage{braket}
\graphicspath{{graphics/}}
\usepackage[hidelinks]{hyperref}
\hypersetup{
    colorlinks,
    citecolor=blue,
    filecolor=black,
    linkcolor=black,
    urlcolor=blue
}

\newcommand{\NNLO}[0]{N$^{2}$LO$_{\rm sat}$}
\newcommand{\EMint}[0]{1.8/2.0 (EM)}

\newcommand{\abinitio}[0]{\textit{ab initio}}

\begin{document}

\author{M.~S.~Martin}\email{matthew\_martin\_3@sfu.ca}
\affiliation{TRIUMF, 4004 Wesbrook Mall, Vancouver, BC V6T 2A3, Canada}
\affiliation{Department of Physics, Colorado School of Mines, Golden, CO 80401, USA}
\affiliation{Department of Physics, Simon Fraser University, Burnaby, BC V5A 3S6, Canada}
\author{S.~R.~Stroberg}
\affiliation{Department of Physics, University of Washington, Seattle, WA 98195, USA}
\author{J.~D.~Holt}
\affiliation{TRIUMF, 4004 Wesbrook Mall, Vancouver, BC V6T 2A3, Canada}
\affiliation{Department of Physics, McGill University, Montr\'eal, QC H3A 2T8, Canada}
\author{K.~G.~Leach}
\affiliation{Department of Physics, Colorado School of Mines, Golden, CO 80401, USA}

%%%%%%%%%%%%%%%%%%%%%%%%%%%%%%%%
\title{Testing isospin symmetry breaking in ab initio nuclear theory}

\begin{abstract}
In this work we present the first steps towards benchmarking isospin symmetry breaking in \abinitio{} nuclear theory for calculations of superallowed Fermi $\beta$-decay.
Using the valence-space in-medium similarity renormalization group, we calculate $b$ and $c$ coefficients of the isobaric multiplet mass equation, starting from two different Hamiltonians constructed from chiral effective field theory.
We compare results to experimental measurements for all $T=1$ isobaric analogue triplets of relevance to superallowed $\beta$-decay for masses $A=10$ to $A=74$ and find an overall agreement within approximately 250~keV of experimental data for both $b$ and $c$ coefficients.
A greater level of accuracy, however, is obtained by a phenomenological Skyrme interaction or a classical charged-sphere estimate.
Finally, we show that evolution of the valence-space operator does not meaningfully improve the quality of the coefficients with respect to experimental data, which indicates that higher-order many-body effects are likely not responsible for the observed discrepancies.
\end{abstract}
\maketitle

%%%%%%%%%%%%%%%%%%%%%%%%%%%%%%

\section{Introduction}
\subsection{Fundamental Symmetry Tests}

Precision measurements of superallowed $0^{+}\to0^{+}$ $\beta$-decays are a critical tool to search for physics beyond the Standard Model in the quark sector~\cite{Hardy2020}.  
This is possible because the decay mode is independent of any axial-vector contribution (up to radiative corrections), and thus provides the most stringent determination of the vector coupling strength in the weak interaction, $G_V$~\cite{Patrignani2016}.  
In fact, the up-down element of the Cabibbo-Kobayashi-Maskawa (CKM) quark-mixing matrix, $V_{\mathrm{ud}}$, is the most precisely known (to the level of $0.032\%$) and relies nearly entirely on superallowed $\beta$-decay $ft$-values determined from measurements of the half-life, $Q$-value, and branching fraction of the superallowed mode~\cite{Hardy2020}.

In order to use the experimental superallowed data to test the Standard Model, small corrections to the $\beta$-decay $ft$-values must first be made to obtain nucleus-independent ${\cal F}t$ values,
\begin{eqnarray}
{\cal F}t\equiv ft(1+\delta_R)(1-\delta_C)=\frac{2\pi^3\hbar^7\ln(2)}{2G_V^2m_e^5c^4(1+\Delta_R)},
\label{Ft_value}
\end{eqnarray}
where $\delta_R$ is a transition-dependent radiative correction, $\Delta_R$ is a transition-independent radiative correction, and $\delta_C$ is a nucleus-dependent isospin-symmetry-breaking (ISB) correction. 
Although these values are relatively small (typically $1\%$ or less), the precision of the experimental $ft$-values is so good ($\leq0.1\%$) that it is critical to take these theoretical corrections into account. 
In fact, the overall uncertainty of $G_V$, and consequently $V_{\mathrm{ud}}$, is currently dominated by $\Delta_R$ and $\delta_C$~\cite{Hardy2020}.
Interest in the theoretical corrections has grown dramatically after a re-evaluation of $\Delta_R$~\cite{Seng2018,Seng2019,Czarnecki2019} led to a significant deviation from the top-row sum unitarity condition of the CKM matrix.
Additionally, the leading contribution to the uncertainty in $V_{\rm ud}$ is now due to the ISB correction from nuclear structure theory~\cite{Hardy2020}.
Therefore, efforts to improve the analysis of these uncertainties within a given theoretical framework is now perhaps one of the most important aspects in this field.

The current extraction of $V_{\mathrm{ud}}$ from the superallowed data uses the shell-model ISB corrections of Towner and Hardy (TH), largely because of the impressive experimental testing to which their formalism has been exposed~\cite{Hardy2020}. 
One lingering issue, however, is that the phenomenological character of these calculations makes it unclear how to robustly quantify their uncertainties or systematically improve them in a controlled manner~\cite{Towner2010,Miller2009}.
Despite recent progress in adapting theoretical methods for calculating ISB corrections relevant for superallowed $\beta$-decay \cite{Satula2012,Satula2016,Xayavong2018,LeachHolt18}, capturing meaningful uncertainties in the quoted errors for the ISB corrections on a case-by-case basis remains a significant challenge.
In this article we therefore present the first steps towards understating the details of these calculations from the \abinitio{} valence-space in-medium similarity renormalization group (VS-IMSRG), which can consistently cover the range of superallowed systems of interest for fundamental symmetry tests.

%%%%%%%%%%%%%%%
\subsection{Isobaric Multiplet Mass Equation}
Since $\delta_C$ is a purely theoretical quantity in the sense that there is no way to extract it directly from experimental measurements, observables sensitive to ISB effects should first be examined before any firm statements on the quality of $\delta_C$ are made.
One such approach is through the isobaric multiplet mass equation (IMME), which has historically been used to predict binding energies of missing elements of isobaric analogue states (IAS)~\cite{Yazidjian2007,Zhang2012}.
This quadratic equation is obtained by assuming the ISB part of the Hamiltonian is at most a rank-2 spherical tensor in isospin space and evaluating it in first order perturbation theory (see e.g. \cite{Lam2013}). It can then be used in to remove systematic errors due to the much larger isospin-conserving part of the nuclear Hamiltonian to isolate the ISB contributions.
The IMME, when written in terms of mass excesses $M$, is typically expressed as:
\begin{equation}
\label{eq: IMME}
M(\alpha,T,T_z) = a(\alpha,T) + b(\alpha,T)T_z + c(\alpha,T)T_z^2,
\end{equation}
where $\alpha$ is a placeholder for quantum numbers of the state, $T$ is the total isospin of the nucleus, $T_z$ is the total isospin projection of the nucleus, and $a,b,$ and $c$ are fitting coefficients.
For an isospin triplet with $T=1$, the fit is trivial and the $a$, $b$ and $c$ coefficients are directly related to the masses
\begin{subequations}\label{eq:bc}
\begin{align}
a&=M_0 \\
b&=\tfrac{1}{2}(M_{+1}-M_{-1})\\
c&=\tfrac{1}{2}(M_{+1}+M_{-1}-2M_{0})
\end{align}
\end{subequations}
where $M_{T_z}$ is shorthand for $M(\alpha,T=1,T_z)$.

%%%%%%%%%%%%%%%%%%%%%%%%%%%%%%%%%%
\section{Theoretical Methods}

Advances in chiral effective field theory ($\chi$EFT) \cite{Epelbaum2009, Machleidt2011} and similarity renormalization group (SRG) \cite{Bogner2007,Bogner2010} as well as \abinitio{} many-body methods \cite{Barrett2013,Carlson2015,Hagen2014,Hergert2016, Dickhoff2004,Stroberg2019} have enabled converged calculations of essentially all nuclei to $N,Z\sim50$~\cite{Stroberg2017,Morris2018,Stroberg2021}. 
Specifically, the consistent inclusion of three-nucleon (3N) forces in chiral Hamiltonians has improved the accuracy of \abinitio{} methods in the medium-mass region~\cite{Otsuka2010,Hergert2013,Hebeler2015} to the point that they are comparable to phenomenological methods for both ground and excited-state energies \cite{Stroberg2016}. 
However, it has yet to be determined whether these improvements are sufficient to be relevant for superallowed $\beta$-decay \cite{Reiter2017}.

There have been several previous attempts to examine the ability of non-empirical approaches to reproduce experimentally extracted IMME coefficients.
These studies typically used many-body perturbation theory to generate effective valence-space Hamiltonians starting from either NN+3N forces in selected $sd$-shell multiplets~\cite{Holt13PR,Gallant2014,Brodeur2017}, or from various NN-only interactions in the $p$- or $pf$-shells~\cite{Caurier2002,Ormand2017}.
Observed deficiencies in these studies were attributed to either neglected 3N forces or unclear perturbative convergence. Therefore we aim to determine whether a nonperturbative many-body approach with NN+3N forces can potentially improve this picture.

In this work we use two sets of NN+3N forces derived from chiral effective field theory:~\EMint{} from a family of interactions constructed in Ref.~\cite{Hebeler2011} and \NNLO{} \cite{Ekstrom2015}.
These interactions were chosen because one (\EMint) has been shown to systematically reproduce ground-state energies to the tin region~\cite{Simonis2016,Hagen2016,Morris2018}, while the other (\NNLO) accurately reproduces absolute and relative nuclear charge radii~\cite{Lapoux2016,GarciaRuiz2016,Simonis2017,Groo20Cu}.
We solve the many-body problem via the VS-IMSRG method~\cite{Tsuk12SM,Hergert2016,Stroberg2019,IMSRGCode}, where an approximate unitary transformation is derived to decouple a given core energy in addition to an effective valence-space Hamiltonian. We subsequently diagonalize using the code NuShellX@MSU~\cite{Brown2014} to obtain absolute binding energies for all members of the $T=1, J^{\pi}=0^{+}$ IATs for mass numbers $A=10$ through $A=74$.

We work at the IMSRG(2) approximation, in which we normal order all operators with respect to finite-density reference state $|\Phi\rangle$, and discard all residual 3N operators, including those induced by the IMSRG evolution.
This approximation has been found to be accurate for absolute ground state energies at the level of a few percent~\cite{Roth2012,Hergert2013,Stroberg2017}.
Naively, this should lead to errors of a few MeV for the IMME $b$ and $c$ coefficients, which would preclude a meaningful comparison with experiment.
However, the error due to the truncation is highly correlated between members of an IAT, and cancels to a significant extent when taking differences \cite{Martin_Thesis} (see also~\cite{Stroberg2021}).

An exception to this occurs if different valence spaces are used for different members of the IAT.
For example, the $A=18$ IAT includes $^{18}$O which has a closed shell for protons, so the natural valence space would only involve neutrons in the $sd$-shell.
Decoupling the proton $sd$-shell involves additional transformations which, given the IMSRG(2) approximation, deteriorates the accuracy of the absolute ground state energy.
On the other hand, $^{18}$F requires an active valence space for both protons and neutrons.
Our present interest is in the IMME coefficients rather than absolute binding energies, and so for all calculations in this work, we use a consistent valence space for each member of an IAT.
The impact of this choice was investigated further in Ref.~\cite{Martin_Thesis}.

An additional ambiguity arises due the choice of reference state $|\Phi\rangle$ for performing the normal ordering.
The usual prescription we have followed in past work is to use a spherically symmetric ensemble reference with average proton and neutron numbers corresponding to the target nucleus (see the discussion in Ref.~\cite{Stroberg2019}).
This captures bulk effect of 3N interactions (both input and induced) between valence particles, but due to the IMSRG(2) approximation, some contributions are missed.
If we explicitly retained all many-body operators induced during the IMSRG flow, the result would be independent of the choice of reference.
Therefore, exploring different reference choices provides a handle on the IMSRG(2) truncation error.
In this work, we follow four different prescriptions: (i) compute each member of the triplet with their own reference, or compute all three members using the same reference, which can be that of either the (ii) $T_z=1$, (iii) $T_z=0$, or (iv) $T_z=-1$ member.

For all calculations, we work in a basis built from harmonic-oscillator states up to a cutoff ($e=2n+\ell \leq e_{\mathrm{max}}$) with $e_{\mathrm{max}}=6,8,10,12$.
For $N=Z$ nuclei where the ground state does not have $J^{\pi}=0^{+}$, the lowest excited state with this spin-parity configuration is used to complete the multiplet.
From there, binding energies are converted to mass excesses for each nucleus and the $e_{\mathrm{max}}=8,10,12$ points fit to an exponential and extrapolated to $e_{\mathrm{max}}\to\infty$.
Finally, the $b$ and $c$ coefficients are obtained from the extrapolated masses via Eq.~\ref{eq:bc}.

\begin{figure}[t]
\includegraphics[width=\linewidth]{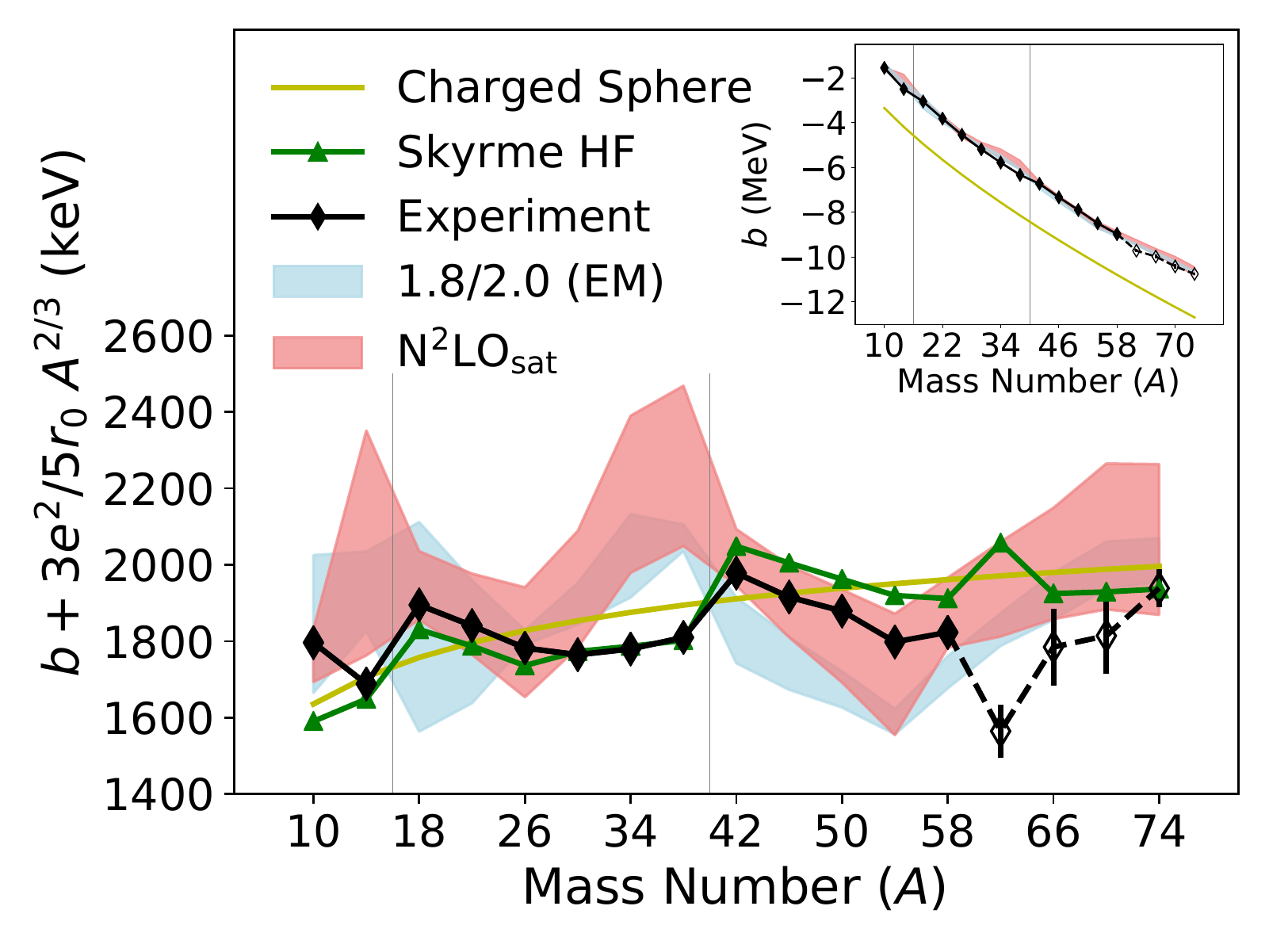}
\caption{\label{fig: b plot} (Colour online) IMME $b$ coefficients both absolute (insert) and with contributions from a sphere of radius $R=r_0 A^{1/3}$, $r_0=1.2$~fm subtracted. The shaded bands for the VS-IMSRG calculations indicate sensitivity to the normal-ordering reference. Charged sphere calculations are for a uniform (insert) and Woods-Saxon (main) charge distribution.}
\end{figure}

\section{IMME results}
\subsection{Effects of Normal-Ordering Reference State}

The resulting IMME $b$ coefficients are plotted in Fig.~\ref{fig: b plot} as a function of mass number $A$, compared with experimental data (note that the experimental points for $A\geq 62$ rely on extrapolated masses).
As there is no \textit{a priori} reason to favour any given reference state, and none of the four are substantially better or worse at predicting experimental data, VS-IMSRG calculations are shown as a band rather than four distinct curves.

To focus on the non-trivial structure, we subtract the charged sphere value $b=-\frac{3e^2}{5r_0}A^{2/3}$, with $r_0=1.2$~fm.
For comparison, we show the value obtained from the potential energy of a classical charged sphere with a Woods-Saxon charge density profile with radius parameter $R=r_0A^{1/3}$~fm and diffuseness $a=0.524$~fm \cite{Krane1987}. We also show the result of a spherical Skyrme Hartree-Fock (Skyrme) calculation using the SKX interaction~\cite{Brown1998,Brown2000,Brown2002}, performed with the code \texttt{dens}~\cite{Brown_Private_Dens}.
We note the SKX interaction was fit to the binding energies of several closed-shell nuclei, with the Coulomb exchange term turned off to better reproduce the $^{48}$Ca-$^{48}$Ni binding energy difference.
For open-shell nuclei, we use the equal filling approximation in the Skyrme calculations.

\begin{figure}
\includegraphics[width=\linewidth]{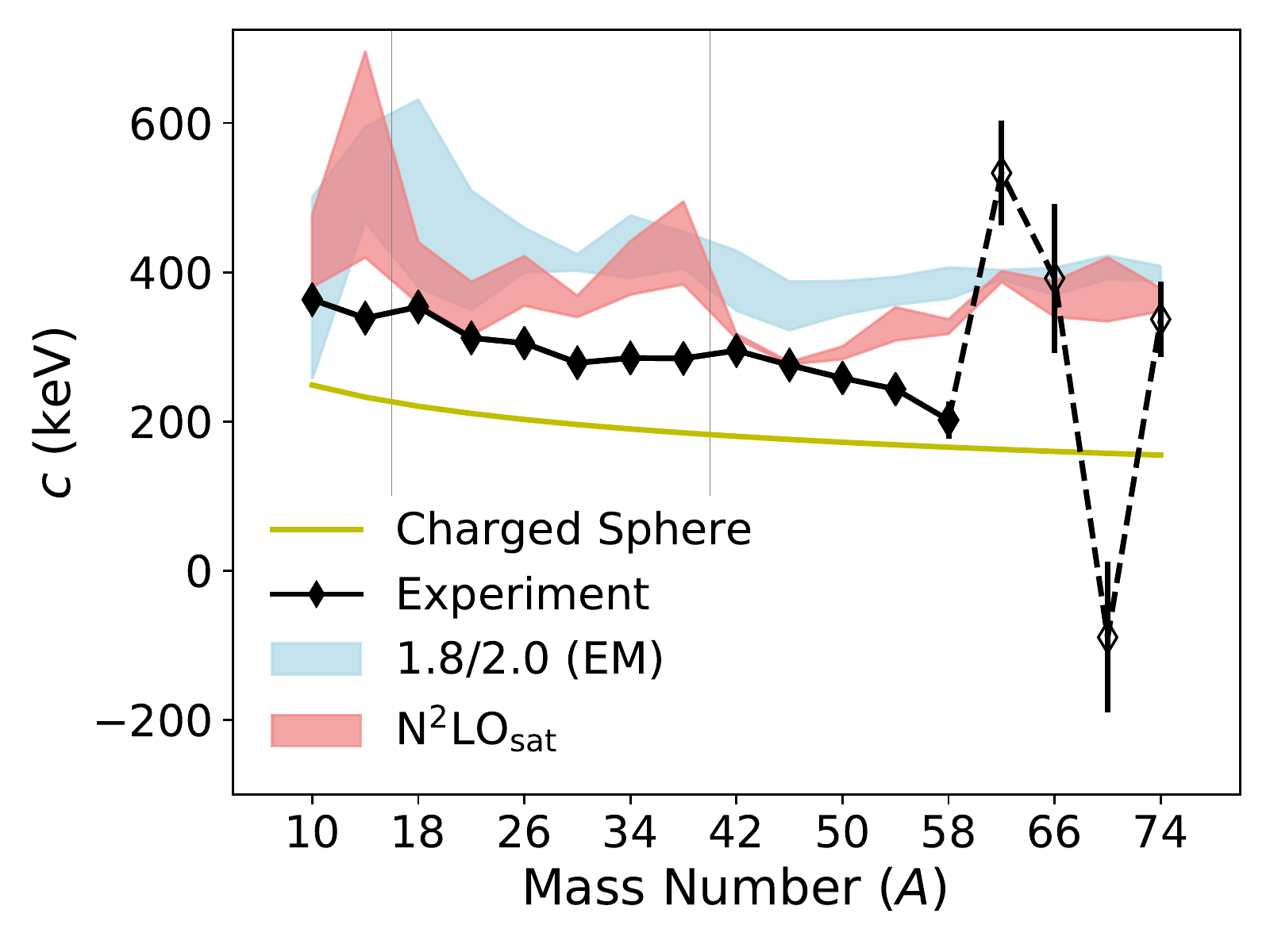}
\caption{\label{fig: c plot} (Colour online) IMME $c$ coefficients. The shaded bands for the VS-IMSRG calculations indicate sensitivity to the normal-ordering reference. Charged sphere calculations are for a Woods-Saxon charge distribution.}
\end{figure}

We see that the general trend follows the simple Woods-Saxon sphere prediction, while finer details are reproduced by the the Skyrme calculation. This indicates that details emerge at the mean field level without explicit treatment of pairing or deformation effects.
The $b$ coefficients obtained with the VS-IMSRG are largely consistent with experiment within the uncertainties of the reference dependence, with a notable deviation in the upper $sd$-shell at $A=34, 38$.
A similar deviation occurs in the upper $p$-shell at $A=14$, though it is somewhat washed out by the reference dependence.
We expect that this deviation is due to IMSRG(2) truncation errors, which are enhanced near the limits of the valence space, and not captured through variation of the reference.
This expectation is supported by the observation that similar deviations are obtained with both chiral interactions; this issue is explored further in Sec.~\ref{Sec: IMSRG evolution}.

The fact that the results are largely interaction independent is somewhat surprising.
The \EMint{} interaction generally predicts charge radii $\sim$ 3-5\%~\cite{Simonis2017} below experiment.
Given the $1/R$ dependence of the $b$ coefficient, this trend in the radii should correspond to an increase on the order of a few hundred keV that further grows with mass.
Comparing calculations from \EMint{} and \NNLO{} does not, however, reflect this expectation.
This can potentially be understood by considering that the small radii and resulting greater Coulomb repulsion means proton orbits are pushed further outward, reducing their kinetic energy relative to neutron orbits and partially cancelling the effect.

%%%%%%%%%%%%%%%%%%%%%%%%%%%%%%%%%%%
%%% Start of C coefficients
%%%%%%%%%%%%%%%%%%%%%%%%%%%%%%%%%%%
Since the IMME $c$ coefficients vary slowly with mass number (the charged-sphere estimate is $c=\frac{3e^2}{5r_0}A^{-1/3}$), we plot them directly in Fig.~\ref{fig: c plot}.
The experimental values are compared with the results of the VS-IMSRG calculations and the estimate from a classical charged sphere with the same Woods-Saxon density profile as used for the $b$ coefficient.
Because the Skyrme calculation only yields ground state energies and shell model USD calculations (e.g. \cite{Magilligan2020}) fit to the IMME coefficients, no meaningful comparisons to phenomenological methods are made for the $c$.

Here we see the Woods-Saxon sphere estimate lies systematically below the data, while
the VS-IMSRG results lie systematically above, with the reference dependence on the same order or slightly smaller than the deviation.
Similar to the $b$ coefficients in Fig.~\ref{fig: b plot}, the VS-IMSRG values near harmonic oscillator shell closures in Fig.~\ref{fig: c plot} show an increased deviation from experiment compared to mid-shell.
While we again expect this is due to the IMSRG(2) approximation, the impact on the $c$ coefficient appears less pronounced.
With the $c$ coefficient also having a $1/R$ dependence, the smaller charge radii predicted by the \EMint{} should lead to an increase in the $c$ coefficient magnitude on the order of a few percent.
This can be seen in Fig.~\ref{fig: c plot}, but the effect is mostly washed out by the reference dependence for normal ordering.

To estimate the relative contributions of Coulomb and strong ISB forces, we consistently IMSRG evolved the Coulomb operator and evaluated it in first order perturbation theory for the $sd$ shell cases.
We found this accounts for 1/3 to 1/2 of the magnitude of the $c$ coefficient.
The remaining contribution comes both from strong ISB forces and from isospin-conserving forces acting on Coulomb distorted wave functions.

\begin{figure}
\includegraphics[width=\linewidth]{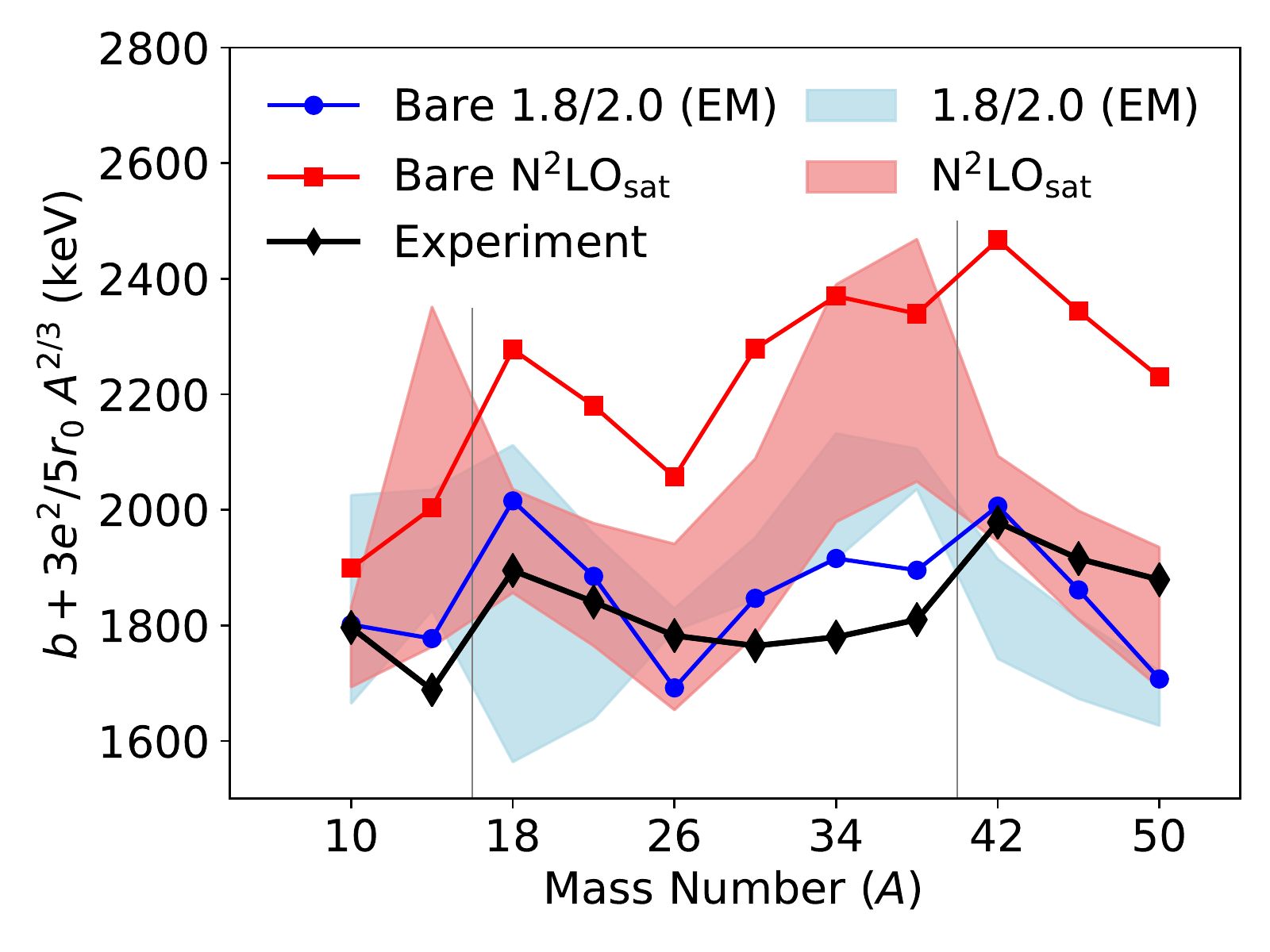}
\caption{(Colour online) IMME $b$ coefficients for \abinitio{} calculations both with and without IMSRG evolution, subtracting the contributions from a uniformly charged sphere of radius $R=1.2A^{1/3}$ fm. \label{fig: NoIMSRG b}}
\end{figure}

%%%%%%%%%%%%%%%%%%%%%%%%%%%%%%%%%%%
%%% Start of IMSRG Effects
%%%%%%%%%%%%%%%%%%%%%%%%%%%%%%%%%%%
\subsection{Effects of IMSRG Evolution} \label{Sec: IMSRG evolution}
The most prominent feature of the VS-IMSRG calculations in Figs. \ref{fig: b plot} and \ref{fig: c plot} are the deviations near harmonic-oscillator shell closures.
However, these inconsistencies are well documented limitations of the IMSRG(2) approximation (e.g. \cite{Stroberg2017}).
With this in mind, it is expected that moving beyond IMSRG(2), and retaining at least some 3N operators, should reduce deviations seen at harmonic-oscillator shell closures.
Unfortunately, this cannot be investigated directly at this time, but we can nevertheless explore the impact of the IMSRG evolution by comparing to calculations with unevolved operators.

\begin{figure}
\begin{center}
\includegraphics[width=\linewidth]{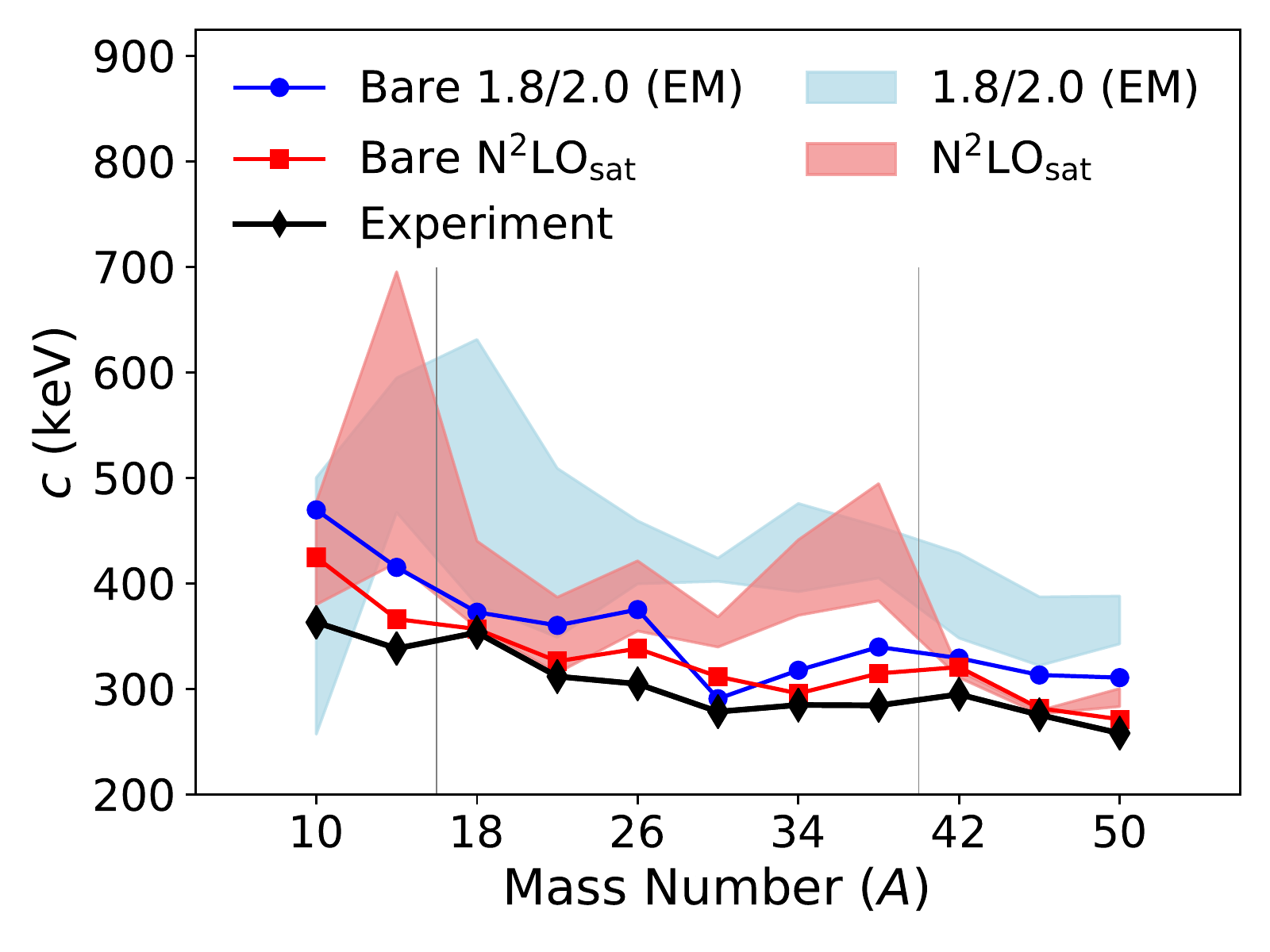}
\caption{(Colour online) IMME $c$ coefficients for \abinitio{} calculations both with and without IMSRG evolution. \label{fig: NoIMSRG c}}
\end{center}
\end{figure}

As illustrated in Ref.~\cite{Martin_Thesis}, the IMSRG evolution acts in roughly the same manner on each mass in an IAT, and as such, the effects are not necessarily obvious.
To further explore this issue, we employ the same two chiral interactions, \EMint{} and \NNLO{}, and calculate IMME coefficients for IATs without performing IMSRG evolution.
These calculations, done at $e_{\mathrm{max}}=12$ and with all operators normal ordered with respect to the Hartree-Fock ground state of the $T_z=0$ nucleus, are compared to the coefficients presented above in Figs.~\ref{fig: NoIMSRG b} and \ref{fig: NoIMSRG c}.
Examining IMME coefficients from the ``bare'' chiral interactions, i.e., those done without IMSRG evolution, shows that deviations near harmonic-oscillator shell closures are generally not present.
While the bare \NNLO{} calculations of the $b$ coefficient do show larger deviations from experimental data than the other cases, they are systematic across all regions.
These observations indicate that the deviations near major oscillator shell closures are indeed a result of the IMSRG evolution.

In comparing to the bare interaction calculations, we further note that there is no apparent improvement from IMSRG evolution.
With the lone exception of the bare \NNLO{} calculations of the $b$ coefficient, better agreement between experimental data and \abinitio{} calculations is always seen for the bare interactions.
This is again somewhat surprising, as absolute ground state energies in nuclei are much better reproduced after IMSRG evolution, and IMME coefficients are directly calculated from binding energies.
Because of the decreased quality of the IMME coefficients after IMSRG evolution, we expect that while moving beyond the IMSRG(2) approximation may help control deviations due to reference state dependence as well as those near harmonic-oscillator shell closures, systematic agreement of calculated IMME coefficients with experimental data may not be improved.

%%%%%%%%%%%%%%%%%%%%%%%%%%%%%%%%%%
\section{Concluding Remarks}
Analysis of the IMME coefficients show that although \abinitio{} calculations are able to systematically reproduce their overall magnitude, the finer details seen in experimental data are generally not.
Dependence on the choice of normal-ordering reference, which would have no effect on the final calculation if all induced operators were retained throughout the IMSRG calculation, are of the same magnitude as both the deviation from experimental data and the dependence on the initial chiral interaction.
Additional deviations when approaching the edge of the employed valence space are observed, and are attributed to the impact of truncating induced many-body forces.
Since IMSRG evolution does not systematically improve agreement with experiment, without a more detailed understanding of the source of this theoretical error, we would not expect moving beyond the IMSRG(2) approximation to substantially improve the reproduction of experimental IMME coefficients.

Although the precise relationship between IMME coefficients and the ISB correction to superallowed $\beta$-decay $ft$-values is not obvious, both depend on the ability to systematically reproduce the effects of ISB in nuclei.
With the recent reduction of uncertainty on radiative correction, the leading uncertainty contribution to the nucleus-independent ${\cal F}t$-value is now that of the ISB correction.
The ability to calculate ISB effects using \abinitio{} methods provides a clear path forward, as these calculations allow for the possibility of a rigorous estimation of theoretical uncertainties, a goal which is challenging, if not impossible, in phenomenological approaches.
With the observed difficulties of \abinitio{} calculations to reproduce of IMME coefficients, parallel improvements in the accuracy of nuclear forces and many-body methods, in addition to a clear blueprint to assess uncertainties, are needed for ISB corrections to be calculated with the level of confidence needed to test physics beyond the Standard Model.

\begin{acknowledgments}
%We would like to thank \commentMSM{SOMEONE?} for enlightening discussions.
We are grateful to J. Simonis and P. Navr\'atil for providing the \EMint{} and \NNLO{} matrix element files. TRIUMF receives funding via a contribution through the National Research Council of Canada. This  work was  supported by the Natural Sciences and Engineering Research Council of Canada (NSERC) and the U.S. Department of Energy (DOE) under contracts DE-FG02-97ER41014, DE-FG02-93ER40789, and DE-SC0017649. Computations were performed with an allocation of computing resources on Cedar at WestGrid and Compute Canada, and on the Oak Cluster at TRIUMF managed by the University of British Columbia department of Advanced Research Computing (ARC).
\end{acknowledgments}

\bibliography{ref}
%\vspace{1in}

%\newpage
\appendix
\section{Tables of IMME coefficients}
We list in Table~\ref{tab:bcoefs} and Table~\ref{tab:ccoefs} the computed $b$ and $c$ coefficients which are plotted in Fig.~\ref{fig: b plot} and Fig.~\ref{fig: c plot}, respectively.

\begin{table} [H]
 \caption{\label{tab:bcoefs}IMME $b$ coefficients calculated using \EMint{} and \NNLO{} interactions. All masses are in keV} 
 \begin{ruledtabular}
 \begin{tabular}{cccccc} 
   & \multicolumn{2}{c}{\EMint{}} & & \multicolumn{2}{c}{\NNLO}  \\ 
 A & min & max & & min & max \\
 \hline
10 & -1677 & -1316 & & -1648 & -1509\\ 
14 & -2356 & -2147 & & -2419 & -1831\\ 
18 & -3381 & -2833 & & -3088 & -2909\\ 
22 & -4015 & -3694 & & -3888 & -3676\\ 
26 & -4528 & -4490 & & -4665 & -4378\\ 
30 & -5107 & -4999 & & -5168 & -4863\\ 
34 & -5641 & -5424 & & -5577 & -5166\\ 
38 & -6102 & -6032 & & -6088 & -5669\\ 
42 & -6957 & -6784 & & -6754 & -6606\\ 
46 & -7570 & -7431 & & -7433 & -7245\\ 
50 & -8145 & -8050 & & -8080 & -7836\\ 
54 & -8729 & -8663 & & -8731 & -8414\\ 
58 & -9111 & -9027 & & -9005 & -8822\\ 
62 & -9490 & -9405 & & -9466 & -9218\\ 
66 & -9902 & -9779 & & -9900 & -9609\\ 
70 & -10291 & -10168 & & -10345 & -9964\\ 
74 & -10736 & -10621 & & -10821 & -10427\\ 
\end{tabular}
 \end{ruledtabular} 
 \end{table}
 
\begin{table} [H]
 \caption{\label{tab:ccoefs}IMME $c$ coefficients calculated using \EMint{} and \NNLO{} interactions. All masses are in keV.} 
 \begin{ruledtabular}
 \begin{tabular}{cccccc} 
    & \multicolumn{2}{c}{\EMint{}} && \multicolumn{2}{c}{\NNLO}  \\ 
 A & min & max && min & max \\
 \hline 
10 & 257 & 500 && 380 & 477\\ 
14 & 467 & 595 && 419 & 695\\ 
18 & 380 & 631 && 358 & 440\\ 
22 & 349 & 509 && 315 & 387\\ 
26 & 399 & 459 && 355 & 421\\ 
30 & 402 & 424 && 340 & 368\\ 
34 & 392 & 476 && 370 & 441\\ 
38 & 405 & 454 && 383 & 494\\ 
42 & 348 & 428 && 311 & 317\\ 
46 & 322 & 387 && 276 & 279\\ 
50 & 342 & 388 && 283 & 300\\ 
54 & 356 & 393 && 308 & 353\\ 
58 & 364 & 406 && 317 & 337\\ 
62 & 390 & 403 && 386 & 401\\ 
66 & 369 & 405 && 340 & 388\\ 
70 & 390 & 422 && 334 & 419\\ 
74 & 387 & 408 && 347 & 378\\ 
 \end{tabular} 
\end{ruledtabular}
 \end{table}

\end{document}